\newcommand{\BibitemShut}[1]{}

\documentclass[iop]{emulateapj-rtx4}
\usepackage{natbib,hyperref,xcolor}

\begin{document}

\title{On-sky performance analysis of the vector Apodizing Phase Plate coronagraph on MagAO/Clio2}

\author{Gilles P.P.L. Otten\altaffilmark{1}, Frans Snik\altaffilmark{1}, Matthew A. Kenworthy\altaffilmark{1}, Christoph U. Keller\altaffilmark{1}, Jared R. Males\altaffilmark{2}, Katie M. Morzinski\altaffilmark{2}, Laird M. Close\altaffilmark{2},  Johanan L. Codona\altaffilmark{2}, Philip M. Hinz\altaffilmark{2}, Kathryn J. Hornburg\altaffilmark{3}, Leandra L. Brickson\altaffilmark{3}, Michael J. Escuti\altaffilmark{3}}

\altaffiltext{1}{Leiden Observatory, Leiden University, P.O. Box 9513, 2300 RA, Leiden, The Netherlands}
\altaffiltext{2}{Steward Observatory, University of Arizona, Tucson, AZ 85721, USA}
\altaffiltext{3}{Department of Electrical and Computer Engineering, North Carolina State University, Raleigh, North Carolina 27606, USA}

\begin{abstract}

We report on the performance of a vector apodizing phase plate coronagraph that operates over a wavelength range of $2-5\,\mu$m and is installed in MagAO/Clio2 at the $6.5\,\mathrm{m}$ \textit{Magellan} Clay telescope at Las Campanas Observatory, Chile. 
The coronagraph manipulates the phase in the pupil to produce three beams yielding two coronagraphic point-spread functions (PSFs) and one faint leakage PSF.
The phase pattern is imposed through the inherently achromatic geometric phase, enabled by liquid crystal technology and polarization techniques.
The coronagraphic optic is manufactured using a direct-write technique for precise control of the liquid crystal pattern, and multitwist retarders for achromatization. 
By integrating a linear phase ramp to the coronagraphic phase pattern, two separated coronagraphic PSFs are created with a single pupil-plane optic, which makes it robust and easy to install in existing telescopes.
The two coronagraphic PSFs contain a 180$^\circ$ dark hole on each side of a star, and these complementary copies of the star are used to correct the seeing halo close to the star. To characterize the coronagraph, we collected a dataset of a bright ($m_L=0-1$) nearby star with $\sim$1.5 hr of observing time.
By rotating and optimally scaling one PSF and subtracting it from the other PSF, we see a contrast improvement by 1.46 magnitudes at $3.5\ \lambda/D$.
With regular angular differential imaging at 3.9 $\mu$m, the MagAO vector apodizing phase plate coronagraph delivers a $5\sigma\ \Delta \mathrm{mag}$ contrast of 8.3  ($=10^{-3.3}$) at 2 $\lambda/D$ and 12.2 ($=10^{-4.8}$) at $3.5\ \lambda/D$.
\end{abstract}

\keywords{instrumentation: high angular resolution, infrared: planetary systems}

\section{Introduction}

In direct imaging, the sensitivity for detecting companions close to the star is primarily limited by residual atmospheric \citep{Racine:99} and quasi-static wavefront variations \citep{marois:05,hinkley:07}. These time-varying wavefront errors manifest themselves as irregularities in the diffraction halo around the star (speckles). 
Coronagraphs reduce the diffraction halo of the star at specific angular scales, and since errors are modulated by diffraction rings, the signal-to-noise ratio (S/N) for companion detection is thus increased.
Both pupil- and focal-plane coronagraphs exist and are used on sky with success \citep{Guyon:06,Mawet:12}.
Many of the latest generation of instruments optimized for high-contrast imaging contain focal-plane coronagraphs, which are typically limited to a raw contrast of $\sim$10$^{-4}$ at small angular separations from the star (a few $\lambda/D$), mostly because of tip/tilt instabilities of the point-spread function (PSF) due to, for example, telescope vibrations and residual seeing effects \citep{Fusco:14,Jovanovic:14,Macintosh:14}.
Pupil-plane coronagraphs are inherently impervious to such effects, as their performance is independent of the position of the star on the science detector, and they can be amplitude- \citep{Carlotti:11} or phase-based \citep{Codona:04}.
One type of pupil-plane coronagraph, called the apodizing phase plate (APP)
coronagraph, is located in the pupil plane and modifies the complex
field of the incoming wavefront by adjusting only the phase
\citep{Codona:06,Kenworthy:07}.
The flux within the PSF of the telescope is redistributed, resulting in a (e.g.,~D-shaped) dark region close to the star.
Since the apodization is with phase only, the throughput of
the APP is higher compared to traditional amplitude apodizers \citep{Carlotti:13SPIE}, and the PSF core only grows slightly in angular size (11.1\% for the phase design in this work).
Because the APP is located in the pupil plane, it is not only
insensitive to residual tip/tilt variations, but also furnishes nodding, chopping, and
dithering motions of the telescope or in the instrument, and indeed observations of close binary stars \citep{Rodigas:15b}.
The PSFs of all stars in the image remain suppressed in the dark hole regardless of the shifts on the focal plane.
In the infrared, the APP can be combined with conventional nodding motions as a thermal background subtraction technique.
Early versions of the APP were realized by diamond-turning a height pattern in a piece of zinc selenide substrate \citep{Kenworthy:07}. 
The phase pattern corresponded to the variation in height of the substrate as a function of position in the telescope pupil (i.e.,~the ``classical phase'' through optical path differences).
As a result of this, the APP was chromatic and suppressed only one side of the star at a time, and the manufacturing was limited to phase solutions with low spatial frequencies.

The vector apodizing phase plate \citep[vAPP,][]{Snik:12} is an improved version of the APP coronagraph and is designed to yield high-contrast performance across a large wavelength range.
In contrast to the regular APP, the phase pattern of the vAPP is encoded in an orientation pattern of the fast axis of a half-wave retarder.
Such a device imposes a positive phase pattern upon right-handed circular polarization and a negative phase pattern upon left-circular polarization, through the geometric (or Pancharatnam-Berry) phase \citep{Pancharatnam:56,Berry:84,Mawet:09}, with the emergent phase pattern equal to $\pm$twice the fast-axis orientation pattern.
This orientation pattern, as well as any other arbitrary pattern, can be embodied by a liquid crystal layer structure, which locally aligns its fast axis to a photo-alignment layer.

The geometric phase is \emph{inherently achromatic}, but leakage terms (which in this case take the shape of the regular PSF) can emerge if the retardance is not exactly half-wave \citep{Mawet:09,Snik:12,Kim:15}.
A typical APP phase design is antisymmetric in the pupil function, which results in a D-shaped dark hole next to the star.
By splitting the circular polarization states with inverse geometric phase signs in the pupil, the vAPP creates two PSFs with dark holes on either side.
By combining multiple self-aligning layers of twisting liquid crystals, it is possible to create retarder structures that have a retardance close to half-wave across a broad wavelength range \citep[up to even more than one octave][]{Komanduri:13}, at wavelength ranges from the ultraviolet (UV) to the thermal infrared (IR).
This class of retarders are called multitwist retarders (MTRs).
The direct-write manufacturing technique of the alignment layer and hence the MTR liquid crystal orientation pattern \citep{Miskiewicz:14} gives high control of the phase of the optic and allows the manufacturing of complex phase designs with typically $\sim 10$ micron spatial resolution that were not
manufacturable using the diamond-turning techniques of earlier APPs.
A vAPP prototype that was optimized for $500-900$ nm was built using both these
techniques, and it was characterized in \citet{otten:14}.

In this paper we present the first on-sky results of the vAPP installed inside the MagAO/Clio2 \citep{Close:2010,Close:13,Sivanandam:2006,Morzinski:2014} instrument on the
$6.5\,\mathrm{m}$ \textit{Magellan}/Clay telescope at Las Campanas Observatory.
We demonstrate the contrast performance at infrared wavelengths at small angular separations from a bright star, and we
show how the two coronagraphic PSFs of the vAPP can be combined to suppress speckle noise inside the dark holes.

\section{The vAPP coronagraph for MagAO/Clio2}
\subsection{The Grating-vAPP principle}
The original implementation of the vAPP included a quarter-wave plate and a Wollaston prism to split circular polarization in a truly broadband fashion. 
Note, however, that leakage terms due to retardance offsets for both the half-wave vAPP optic and the quarter-wave plate limit the contrast performance \citep{Snik:12}.
In \citet{Otten:14spie} we introduced a simplified version of the vAPP (grating-vAPP or gvAPP) that includes a linear phase ramp \citep[i.e.,~a ``polarization grating'';][]{OhEscuti:08,Packham:10} to impose the circular polarization splitting.

For MagAO/Clio2 we have manufactured an infrared version of such a gvAPP device which has a phase pattern that
is composed of two separate patterns: 
the first is an APP phase pattern optimized for the \textit{Magellan} telescope pupil that produces the coronagraphic PSFs with dark D-shaped holes, and the second is a linear phase ramp that is opposite for the two circular polarization states and provides an angular splitting of the two beams with the opposite coronagraphic phase patterns.
This polarization grating splits the two PSFs without the need for a
quarter-wave plate and Wollaston prism, which greatly decreases the cost and enhances the ease of installation.
As both the modification of the PSF and the splitting direction depend on the handedness of circular polarization following the geometric phase, the grating-vAPP produces two separate coronagraphic PSFs with dark holes on opposite sides,
providing continuous coverage around the star.
The inclusion of the linear phase also ensures that the leakage term due
to the plate not being perfectly half-wave ends up between the two
coronagraphic PSFs as a third (unaberrated) PSF.
The positioning of the leakage-term PSF in between the coronagraphic
PSFs minimizes the impact of any residual non-half-wave behavior of the retarder on the contrast inside the dark holes \citep{otten:14}, and thus it enhances the contrast performance with respect to a coronagraph with a quarter-wave plate and Wollaston prism.
This PSF can be used as a photometric and astrometric reference and as an image quality indicator.
Both the structure of the coronagraphic PSFs and their splitting angle are not dependent on
the retardance of the gvAPP device.
Only the brightness ratio of the leakage PSF with respect to the
coronagraphic PSFs changes with varying retardance. 
As the splitting between the coronagraphic PSFs is imposed by a diffractive grating pattern, their separation is a linear function of wavelength.
Hence, while the vAPP optic offers high-contrast coronagraphic performance over a broad wavelength range, to produce sharp PSFs without radial smearing, narrowband filters have to be applied throughout the broad wavelength range over which the device is highly efficient.
By orienting the dark holes left/right with respect to the up/down splitting, this grating effect can furnish low-resolution spectroscopy of point sources inside either of the dark holes. 
Using the gvAPP in combination with an integral field spectrograph overcomes the spectral smearing issue altogether, and such a setup can therefore provide snapshot coronagraphic spectroscopy over the entire efficiency bandwidth.

\subsection{Phase pattern design}

The phase pattern is determined with a simple, iterative algorithm akin to a Gerchberg--Saxton iteration \citep{Gerchberg:72,Fienup:80}. We switch between electric fields in the pupil plane and the focal plane with Fourier transformations and enforce constraints in the corresponding planes. In the pupil plane, the amplitude field amplitude is set to unity inside the telescope aperture and zero everywhere else. In the focal plane, we set the electric field amplitude to zero in the dark hole. This process is repeated hundreds of times until we obtain a phase pattern that achieves the desired contrast. This approach does not guarantee the highest PSF core throughput for a desired contrast, but we found it to perform better than any other design approach that we are aware of.

Since this particular APP design only has a dark hole on one side of the focal plane, the phase pattern in the pupil will be antisymmetric. We use this symmetry to improve the performance of the algorithm. Instead of setting the electric field to zero in the dark hole, we add a scaled and mirrored version to the electric field on the other side of the dark hole. This is motivated by the fact that a one-sided dark hole created by an antisymmetric phase pattern is achieved in the focal plane by symmetric and antisymmetric parts of the electrical field canceling each other in the dark hole and adding to each other on the other side. The scaling enforces energy conservation in the focal plane. A comprehensive description of our design algorithm including applications to symmetric dark holes will be provided in a forthcoming publication by Keller et al.~(2017, in preparation).
For the optimization in this paper, we define a dark hole from 2 to 7 $\lambda/D$ and with a $180^\circ$ opening angle and a desired normalized intensity of $10^{-5}$. The final design has a PSF core throughput of 40.3\% with respect to an unaberrated PSF as the light gets redistributed across the PSF (mostly on the other side from the dark hole).

\subsection{Coronagraph optic specifications}

\begin{figure*}[!ht]
\includegraphics[width=0.5\textwidth]{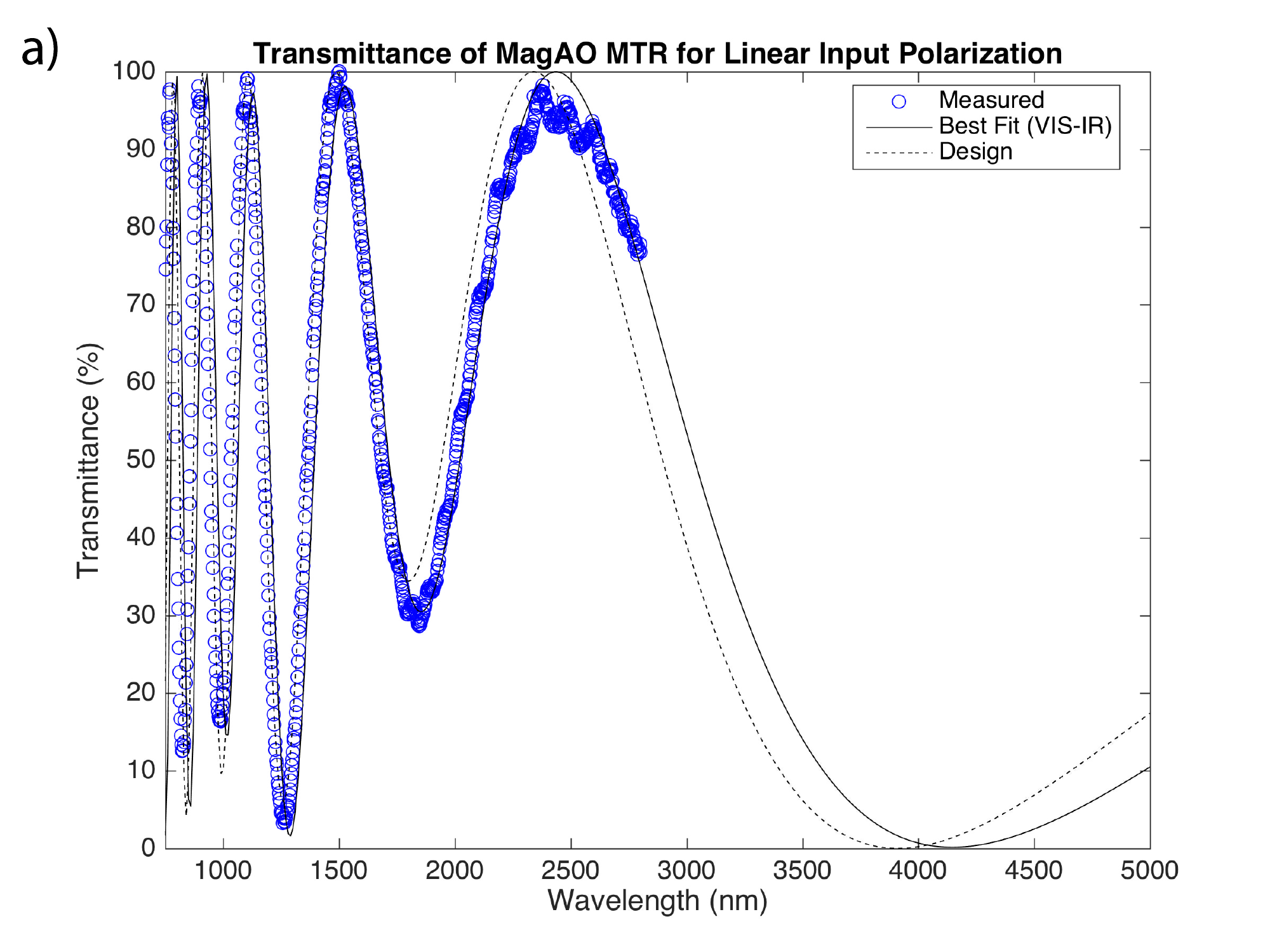}
\includegraphics[width=0.5\textwidth]{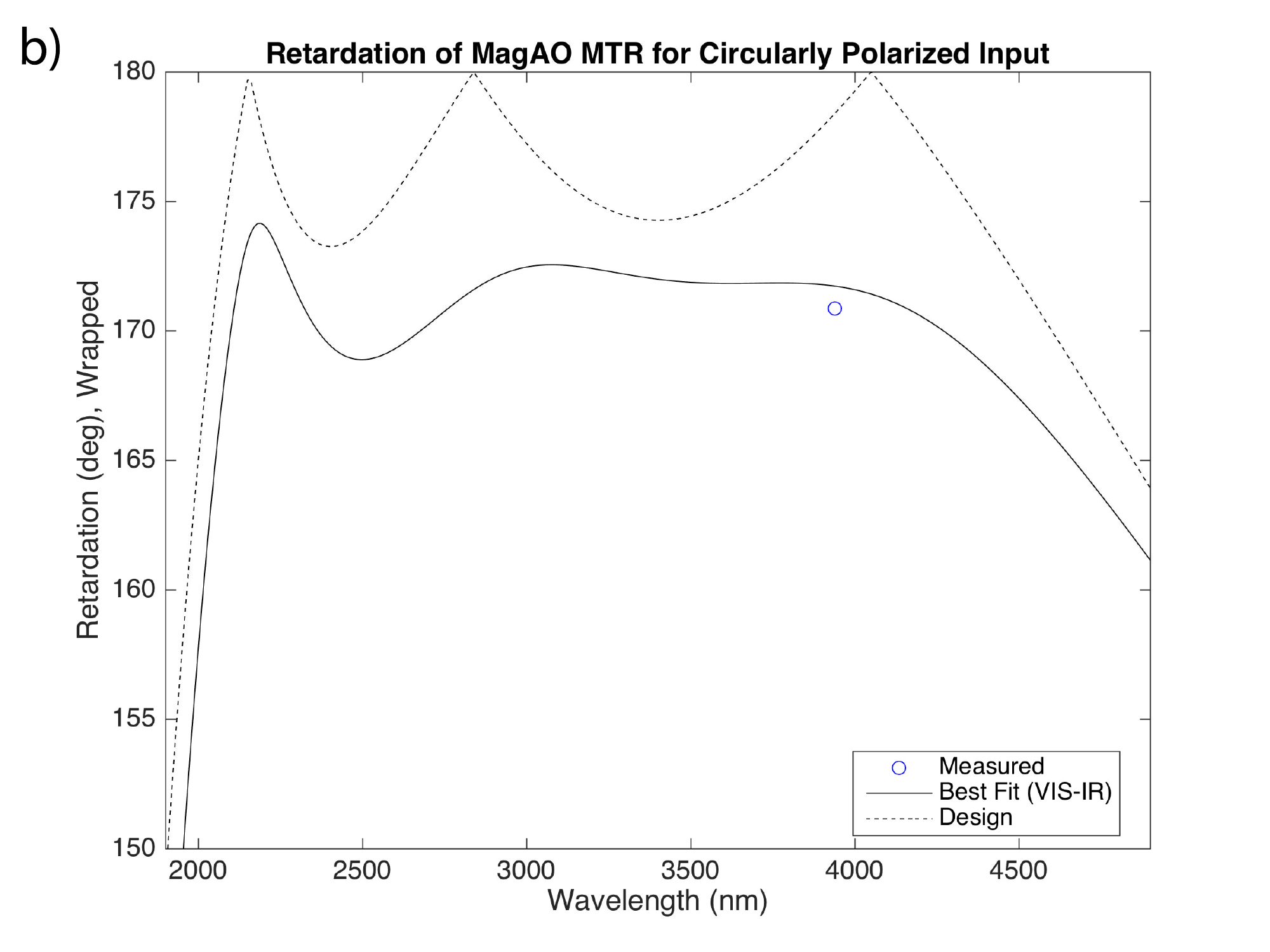}
\includegraphics[width=0.5\textwidth]{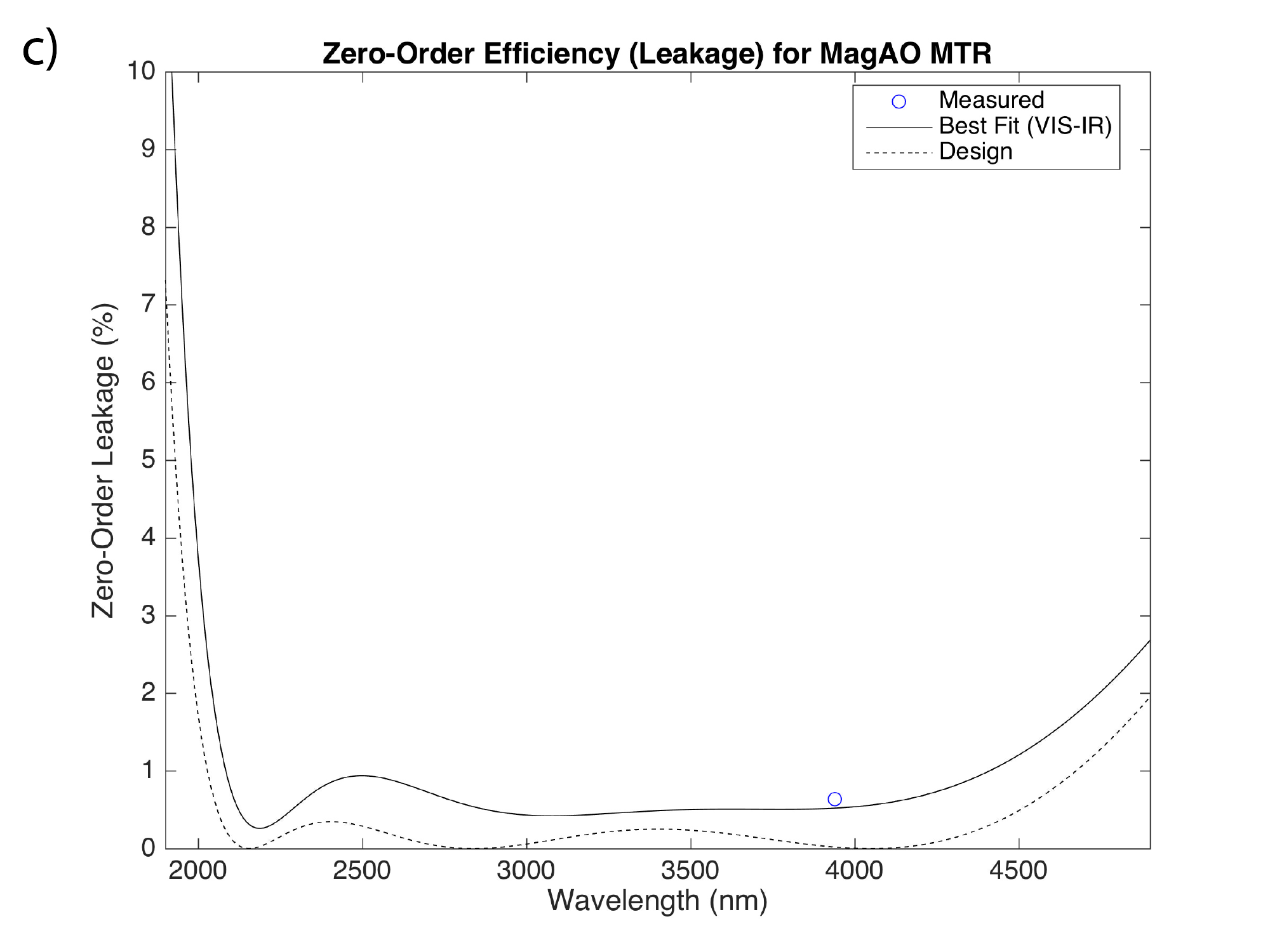}
\caption{(a) Transmission of the vAPP optic between crossed polarizers against wavelength for the theoretical design and a test article with a linear fast axis made according to the same recipe. A model of the MTR is fitted to the test article. (b) Plot of retardance vs. wavelength based on the design and best fit of the MTR model to the crossed polarizer transmission. The retardance requirement corresponds to a maximum leakage of 3.5\% and a retardance offset of 0.38 radians. The on-sky measured datapoint of the leakage is converted into retardance and shown with a blue circle. The measurement error on the datapoint is $0\fdg15$ degrees (estimated by propagating the standard deviation of the leakage to retardance) and is smaller than the blue circle that was used. (c) Percentage of leakage with respect to the total transmitted light corresponding to the wavelength-dependent retardance for both the theoretical design and the best-fitting model of the test article. The on-sky measured datapoint of the leakage is shown with a blue circle.}
\label{fig:retardance}
\end{figure*}

The gvAPP optic has a diameter of $25.4$ mm and a thickness of approximately $3.3$ mm and is designed to work with the Clio2 camera with a nominal size of 3.32 mm of the reimaged \textit{Magellan} telescope pupil.
The diameter of the vAPP pupil mask was undersized by $100$ microns (from a diameter of 3.32 to
3.22 mm) to create a tolerance against pupil misalignments in the instrument.
A $1^\circ$ wedge is added on one side of the coronagraph in order to deflect reflection ghosts.
To further suppress ghost reflections and improve the overall transmission, both sides of the optic are broadband antireflection coated with an average transmission between $2$ and $5$ microns of 98.5\%.
An aluminum aperture mask, matching the \textit{Magellan} pupil, with a pixelated edge (with a pixel size of $11.54$ microns) is deposited on one of the substrates and is sandwiched directly against the retarder layers, manually aligned using a high-power microscope, and fixed in place with an optical adhesive.
The phase pattern (the coronagraphic pupil phase pattern plus the grating pattern) is written as an orientation pattern of an alignment layer of ``DIC LIA-CO01'' by a UV laser with polarization-angle control \citep{Miskiewicz:14}. The pixel size is $11.54$ microns for both the phase and amplitude pattern.
During fabrication, the writing accuracy of the fast axis is calibrated to approximately $2^\circ$, corresponding to a maximum phase error of $4^\circ$, that is, $\sim \lambda/100$.
The patterned retarding layer consists of three MTR layers (Merck RMS09-025; see also Table \ref{tab:tab1}) and is optimized to produce a retardance $\delta$ that is half-wave to within 0.38 radians for wavelengths between $2$ and $5$ microns, corresponding to a maximum flux leakage from the coronagraphic PSFs to the leakage-term PSF of 3.5\%. The design recipe of the MTR is [$\phi_1=78^\circ$, $d_1=3.5 \,\mu$m, $\phi_2=0^\circ$, $d_2=7.3\,\mu$m, $\phi_3=-78^\circ$, $d_3=3.5\,\mu$m], where $d_i$ stands for layer thickness, $\phi_i$ for the twist of a layer, and $i$ for the layer number \citep[see][]{Komanduri:13}. 
This recipe is used to build our coronagraph with our custom fast-axis pattern and also a test article with the same parameters but a fixed fast axis. The transmission of this test article is measured between crossed linear polarizers with a VIS-NIR spectrometer up to 2800 nm. A model of the MTR is fitted to the observed transmission between crossed polarizers with five free parameters (three thicknesses and two relative twists with respect to the middle layer). The best-fit parameters are [$\phi_1=81^\circ$, $d_1=3.5\,\mu$m, $\phi_2=0^\circ$, $d_2=7.3\,\mu$m, $\phi_3=-77^\circ$, $d_3=3.9\,\mu$m], and are used afterward to predict the transmission, retardance, and leakage at wavelengths out to 5000 nm, as shown in Fig.~\ref{fig:retardance}.

The leakage PSF intensity is derived by measuring the peak ratio of either of the coronagraphic PSFs to the leakage-term PSF in a sequence of unsaturated images. The mean and standard deviation of the ratio in this sequence are $31.47\pm1.07$. This ratio is divided by the theoretical PSF core throughput (i.e.,~Strehl) of 0.403 to yield the ratio as if the coronagraph were not present. This means that the intensity of the leakage term is $1/78.1 \cdot I_\mathrm{coron}$, where $I_\mathrm{coron}$ is the intensity of the coronagraphic PSF. 
This value is normalized by the total intensity $(2+1/78.1) \cdot I_\mathrm{coron}$ to yield the fractional leakage intensity (the amount of light that goes into the leakage term). In the completed coronagraph, we measure a leakage-term intensity of 0.636\% at $3.94\, \mathrm{microns}$, which corresponds to $\delta =2.98\,\mathrm{rad}$, using this method, which is within the previously defined specifications. While this leakage is slightly larger than the theoretical expectation at that wavelength (0.16\%), it is comparable in magnitude to the maximum retardance offset of the curve (see Fig.~\ref{fig:retardance}c).

The polarization grating pattern spans $17.5$ waves in terms of phase, corresponding to a displacement of $35\ \lambda/D$ between the two coronagraphic PSFs.
In this way, both of the coronagraphic PSFs fit on the chip at the longest wavelengths ($M'$ band) while minimizing the contribution of the leakage-term diffraction pattern in the dark holes.
The grating creates a splitting angle that is dependent on the
wavelength in terms of pixels of separation, and so the PSFs are
laterally smeared.
For optimal image quality with smearing of at most 1 $\lambda/D$,
the filter FWHM needs to be $\frac{\Delta \lambda}{\lambda} \leq 0.06$.
Due to the optic's broadband efficiency, filters can be used anywhere between $2$ and $5$ microns for coronagraphic imaging.
Note that even outside the specified wavelength range, the coronagraphic performance is never deteriorated by leakage terms, but the coronagraphic PSFs are less efficient as they lose light to the leakage-term PSF.

After installation inside MagAO/Clio2, we collected pupil image measurements with and without the coronagraph at several IR bands
during good sky conditions and with adaptive optics (AO) to obtain accurate on-sky pupil transmission measurements.
We determine the transmission of the optic from the ratio of the pupil intensity with and without the coronagraph.
The theoretical transmission values are detailed in Table
\ref{tab:tab1} per layer and compared to the measured transmission.
Since the measured retardance is close to half-wave (as expected from
the theory), the thickness of the liquid crystal layers cannot deviate significantly from the
theoretical value.
We therefore set their thicknesses to the fitted values for the MTR recipe, which adds up to $14.7\,\mathrm{microns}$.
The absorption properties of the retarder layer were measured in a 900
nanometer thick sample at a wavelength of $4\,\mathrm{microns}$ and extrapolated to
the $14.7\,\mathrm{micron}$ thick layer.
The absorption coefficient derived from this measurement falls on the
high end of the range seen in Fig. 3 of \citet{Packham:10}, who measured the transmission of a similar family of liquid crystals.
The absorption coefficient of the glue layer is derived from the
spectral transmission graph on the Norland Products
website\footnote{\url{https://www.norlandprod.com/adhesives/NOA\%2061.html}}
and the known thickness of their sample. 
The thickness of the glue layer constitutes the largest uncertainty because it was not
measured during the manufacturing process.
Because the other transmission values are well constrained, we let the thickness of
the glue layer vary as a free parameter to match the observed
transmission.
Our derived glue layer thickness of $50\,\mathrm{microns}$ is not unexpected for glass--glass interface bonding.
The breakdown shows that the throughput is primarily limited by the
optical adhesive NOA-61. The absorption features of both the optical adhesive and the retarding layer are related to the vibrational modes of chemical bonds with carbon, such as C--C, C--O, C--N and C--H. 

The gvAPP coronagraph is located in the pupil stop wheel of Clio2 and oriented with the grating splitting angle perpendicular to the arc traveled by the pupil in the pupil wheel. 
The wedge splitting angle was oriented perpendicular to the splitting direction. 
The orientation of the splitting angle corresponds in theory with splitting the PSFs along the short axis of the chip. 
This leaves a large amount of space along the long axis to nod the PSFs along for background subtraction.
From our PSF measurements we see that the orientation of the PSFs on the chip is approximately 26\degr{} rotated away from the preferred orientation. 
This rotation does not interfere with the background subtraction.

\begin{table*}[!ht]
\begin{center}
\caption{Breakdown of the thickness and transmission properties of the different layers of the gvAPP optic installed in MagAO/Clio2.}
\begin{tabular}{l|llcc}

Layers & Material & Thickness & 3.9 micron & 4.7 micron ($M'$) \\
\hline\hline
AR-coating&$\cdots$&$\cdots$& 0.98& 0.99\\
Substrate with $1^\circ$ wedge &CaF$_2$&0.8 mm&0.99&0.99\\
Amplitude mask & evaporated aluminum & 250 nm&$\cdots$&$\cdots$\\
Bonding glue & NOA-61 epoxy &50 $\mu$m& 0.81& 0.81\\
Substrate &CaF$_2$&1 mm& 0.99& 0.99\\
Retarder layers& Merck RMS09-025 & 14.7 $\mu$m&0.85&$\sim0.85$\\
Alignment layer& DIC LIA-CO01 & 50 nm&$\cdots$&$\cdots$\\
Bonding glue & NOA-61 epoxy &50 $\mu$m&0.81&0.81\\
Substrate &CaF$_2$&1 mm&0.99&0.99\\
AR-coating &$\cdots$ &$\cdots$& 0.98&0.99\\
\hline
Theoretical throughput &$\cdots$&$\cdots$&0.53&0.54\\
Measured throughput &$\cdots$&$\cdots$&0.51&0.54\\
\label{tab:tab1}
\end{tabular}
\end{center}

\end{table*}

\section{Observations}

The observations with the vAPP coronagraph at MagAO/Clio2 were taken during 2015 June 6, 07:38:40 - 10:07:34 UT during excellent atmospheric conditions (with only high cirrus clouds).
The filter used for these observations is the $3.9$ micron narrowband filter with a width of 90 nm and a central wavelength of $3.94$ microns. This filter was chosen to take advantage of the extremely high Strehl ratio of the adaptive optics system at longer wavelengths ($>$95\%), and to make sure the radial smearing ($<0.4\ \lambda/D$) interferes only minimally with the interpretation of the PSF suppression in the dark hole.
The plate scale of the detector is 15.85 arcseconds pixel$^{-1}$ \citep{Morzinski:15}.
The target discussed in this paper to assess the contrast performance of the vAPP is an A-type star with an $L'$-band magnitude between 0 and 1.
The star was selected to be bright and without a known companion to explore the limits of the coronagraph's performance. 
Note that the coronagraphic system with the gvAPP at MagAO/Clio is also fully applicable for fainter stars and has been tested on sky down to magnitude-7 targets. 
The performance of the adaptive optics system remains invariably high down to $R=7$ magnitude stars \citep{Close:12}.
A total of 287 data-cubes were taken on sky, each with 20 subframes and an exposure time of 1 s each.
The dataset has a total on-target exposure time of 5740 s.
The derotator was off during the observations, and the observations span a total of $39\fdg45$ of field rotation.
To perform background subtraction, data cubes were recorded with the PSF off the chip, centered approximately 10 arcsec to the left from the nominal center of the science camera array, so that no sources and ghosts are seen on the same part of the chip. 
This background estimation was repeated four times during the sequence.
No flats have been applied to the data, and a sky correction was made using the off-target nods.

\begin{figure*}[!ht]
\center{\includegraphics[width=\textwidth]{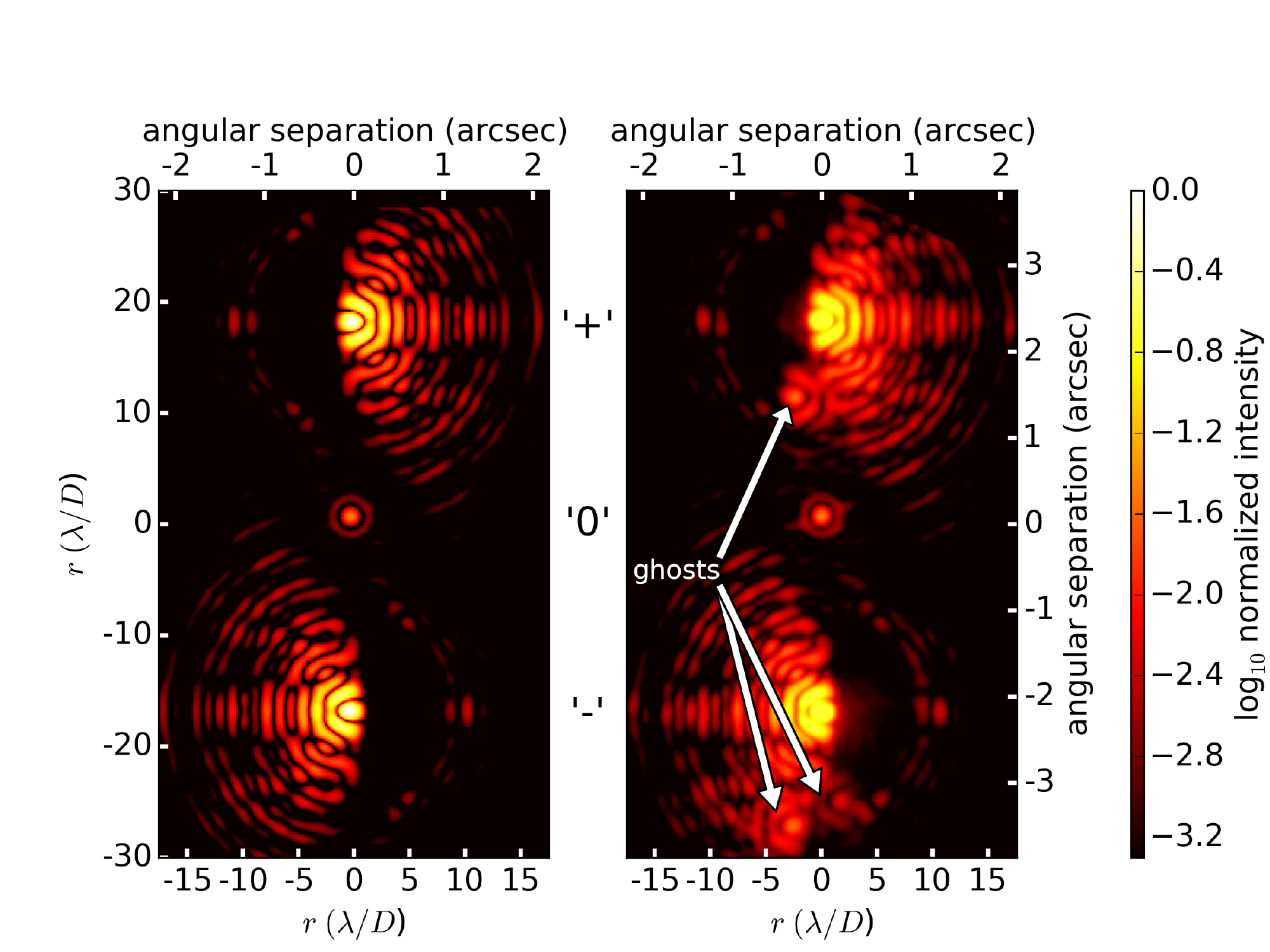}}
\caption{Comparison of theoretical (left) with observed PSFs (right).
Both images are on the same logarithmic scale with a lower threshold of $10^{-3.3}$.
The theoretical PSF is calculated for the central monochromatic wavelength of $3.94$ microns. The retardance in the simulation was set to $\delta = 2.98$ rad, thereby creating  the leakage PSF ``$0$.''
The observed image is saturated on the first diffraction ring and on the central core but is corrected to the unsaturated flux level. A small asymmetry of the intensities of the wind-driven halos of the star inside the dark holes is seen between the two coronagraphic PSFs (``$+$'' and ``$-$''). Instrumental ghosts are indicated with arrows and are not related to the coronagraphic optic. Further investigation showed that these ghosts can be removed from the dark hole by setting the rotator to an angle of $30^\circ$.
}
\label{fig:psfs}
\end{figure*}

Figure~\ref{fig:psfs} shows a comparison between the theoretical and
observed PSFs using a median combination of the top 50\% best frames in
terms of the fitted radius ($1.22\ \lambda/D$) of the leakage-term PSF, acting as a proxy for subframe quality as it expands with increased turbulence.
The observed coronagraphic PSFs are saturated in the core and in the first diffraction ring but are corrected to the peak flux consistent with the unsaturated calibration images.
While the two PSFs have approximately the same brightness, the two PSF halos inside the dark holes have a slightly different intensity. 
A potential source of the difference is discussed in Section \ref{sec:diff}.

\subsection{Data reduction}

To remove hot, dead, or flaky pixels in each image, we subtracted a median-filtered image with a 3$\times$5 pixel box from the cleaned
and centroided image cube to generate an image where the outliers clearly stand out. 
The $3 \times 5$ box is chosen because the outlying pixels tend to have structure in the direction of the readout and not perpendicular to it.
The data points that deviate more than 1000 counts are replaced by the
local value of the 3$\times$5 median.
Four sets of sky reference frames are taken between on-target observations.
The sky frames were median-combined for every one of the four sets.
Each of the previously taken frames on target were subtracted by
the first consecutive median-combined master sky frame.
After the background subtraction, the median of a cosmetically clean
part of the chip is subtracted in order to remove any residual
background offset.
A theoretical diffraction pattern consistent with the geometry of the telescope
and wavelength of observation is used as a fiducial. 
This theoretical PSF is then fit to the central leakage PSF by
minimizing the chi-squared residuals between the theoretical PSF and
the leakage PSF, with $x$, $y$, radius ($1.22\ \lambda/D$), and intensity as the free
parameters of the fit.
All images of the data cubes are coregistered by shifting the images to the
central pixel of the frame with the previously fit $x$ and $y$ values
using a bilinear interpolator.
The radius of the leakage PSF fit is used as a subframe quality indicator.
The few images (39) that have fitted radii significantly smaller ($< 9$ pixels) or larger ($>12$ pixels) than the diffraction limit are excluded.
The best 91\% (5200) of the images are
selected after sorting the frames by radius from smallest to largest.
Both criteria remove the frames
where the seeing conditions temporarily worsened or where the AO system
lost its lock.
After these selections, the images are reordered to
original chronological order and binned by four frames,
corresponding to 4 s integration time per binned frame to reduce memory consumption and computational time.

Instrumental ghosts due to internal reflection of the refractive optics
are present in the image; see Fig.~\ref{fig:psfs}.
Several of these ghosts are typically $10^{-2}-10^{-3}$ in
intensity, and their position relative to the central PSF changes as a
function of position on the chip.
To reduce their influence, the regions of identified ghosts are masked
off from any subsequent fitting or stacking process.
These ghosts can be removed from the dark holes by setting the rotator to an angle of $30^\circ$.

\subsection{Rotation, Scaling, and Subtraction}

\begin{figure*}[!ht]
\center{\includegraphics[width=\textwidth]{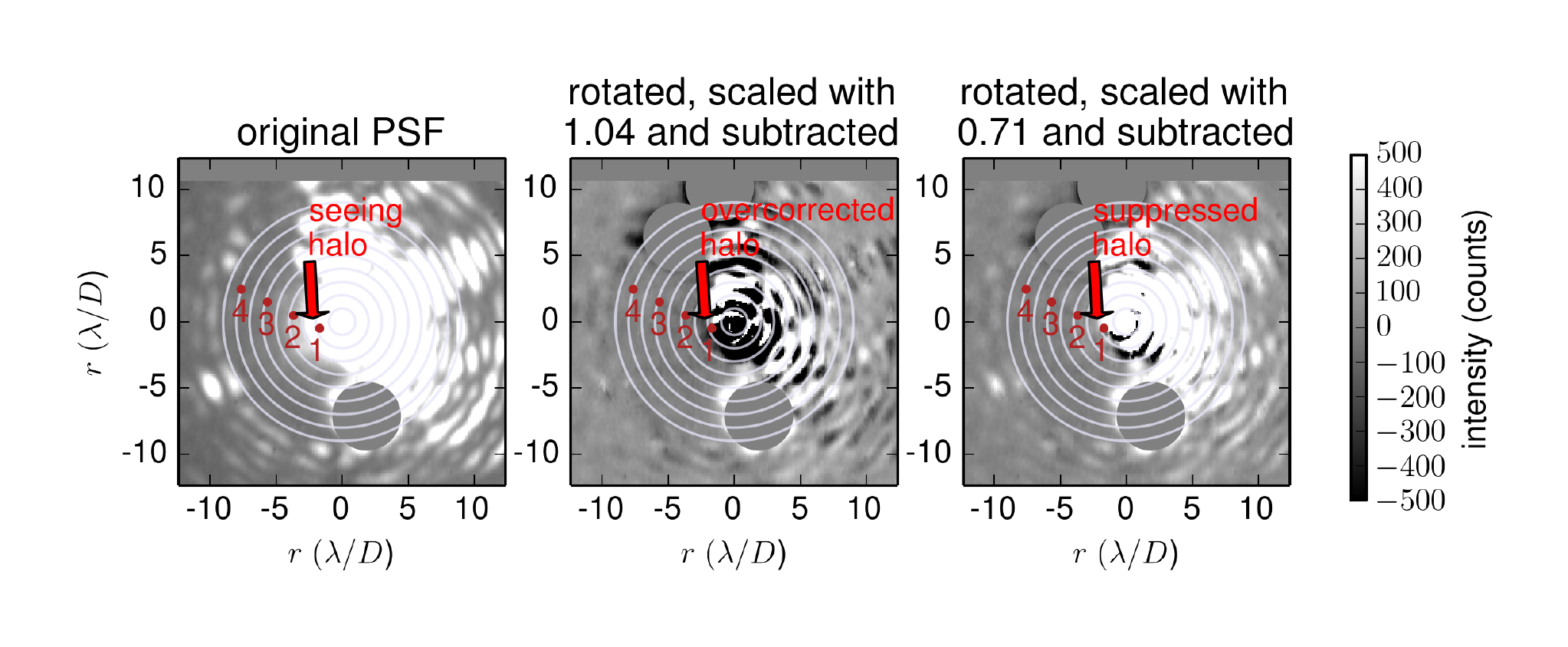}}
\caption{Comparison between subtractions of ``$+$'' and ``$-$'' PSFs for
different scaling factors.
The circles indicate distances from the center in integer $\lambda/D$.
The dark red points and numbers show probe locations that are used for further analysis.
The factor of 1.04 in the rotation-subtraction reduction minimizes the residual diffraction structure, while the factor of 0.71 minimizes the standard deviation across time close to the star.
Note that a companion inside the dark hole in the top PSF would show up as a positive signal, while a detection in the other PSF's dark hole would yield a negative signal.
An instrumental ghost was masked off on both sides of the PSF.
}
\label{fig:drieluik}
\end{figure*}

Each on-target image consists of three PSFs, which we label ``$+$'' for
the upper coronagraphic APP PSF, ``$-$'' for the lower coronagraphic PSF with the dark region on the opposite side of the star, and ``$0$'' for the leakage PSF, which is consistent with the PSF obtained with no coronagraph in the optical path and with a flux typically $10^{-2}$ of the other two PSF cores.
To suppress the noise contribution of the seeing-driven halo inside the dark holes, we use one coronagraphic PSF as a reference for the other coronagraphic PSF of the same star, and we subtract ``$-$'' from ``$+$.''
This PSF subtraction technique avoids self-subtraction of the flux of a potential companion as it is very unlikely to have another companion at the same separation and brightness on the opposite side of the star. 
A similar approach is taken by \citet{Marois:07} and \citet{Dou:15}, who use the (noncoronagraphic) PSF (itself) under rotation as a reference and measure an order-of-magnitude improvement compared to regular LOCI \citep{lafreniere:07} without rotation.
Our approach works by rotating, scaling, and subtracting PSF ``$-$'' from PSF ``$+$'' in a three-step process. 
First, the image is flipped in both dimensions so that ``$-$'' has the same orientation as ``$+$''.
We align each ``$+$'' and ``$-$'' PSF with the median of all ``$+$'' PSFs, by performing a cross-correlation on a bright, isolated feature at $10\ \lambda/D$ on the bright side of the PSF.
With the obtained centroids, the ``$+$'' and ``$-$'' PSFs are subpixel shifted to
the frame center with the python routine \texttt{scipy.ndimage.interpolation.shift} set to first-order spline interpolation.
The ``$-$'' cube is then multiplied by a fixed amplitude ratio and subtracted from the ``$+$'' cube.
An example of the PSFs before and after subtraction can be seen in
Fig. \ref{fig:drieluik} for three different scaling factors (0, 1.04, and 0.71).
The diffraction structures on the bright side of the PSFs are optimally canceled using an intensity
scaling ratio of 1.04.
This is consistent with the ratio of the encircled energies of both PSFs.
However, with this ratio, the seeing-driven halo in ``$-$'' is
oversubtracting the halo in ``+,'' resulting in larger amounts of speckle noise
in the final combined image.
A likely cause for this is that aberrations create pinned speckles on the diffraction structure in the dark holes, but the intensities may be different in the left and right dark holes.
Although this diffraction structure ideally has an intensity of $<10^{-5}$ with respect to the PSF core (and therefore is not visible in the left panel of Fig.~\ref{fig:psfs}), it becomes brighter due to residual seeing or quasi-static aberrations of the telescope and instrument.
Because this diffraction structure is fully point-symmetric between the two PSFs, a rotation-subtraction approach with a variable scaling factor always reduces the pinned speckle structure in the halo.
A simple simulation shows that with the realistic seeing and AO performance the intensity of this halo is practically balanced, even when an AO loop time lag (3 ms) and strong wind speed ($10\,\mathrm{m\,s}^{-1}$) in the worst-case direction of the dark hole orientations are taken into account.
As for (quasi-)static optical aberrations, only odd modes (like trefoil aberration) cause an asymmetry between the holes in the two dark holes, while even modes generate complete symmetric PSF structures. 
However, to first order, odd aberrations will also merely brighten the symmetric diffraction structure inside the dark holes, just with different intensities.
Assuming trefoil is the dominant aberration, we simulate how much trefoil could create a PSF that is still consistent with the observed in terms of the asymmetry between the dark holes. 
Based on this simulation, we conclude that the RMS error of the trefoil aberration needs to be $\sim$0.04 radians (25 nm at 3.94 microns) to match the observations. 
We therefore conclude that the vAPP is not only insensitive to tip/tilt errors, but, through the rotation-scaling-subtraction technique, can also generically cope with low-order wavefront errors.\label{sec:diff}

\begin{figure}[!ht]
\center{\includegraphics[width=0.45\textwidth]{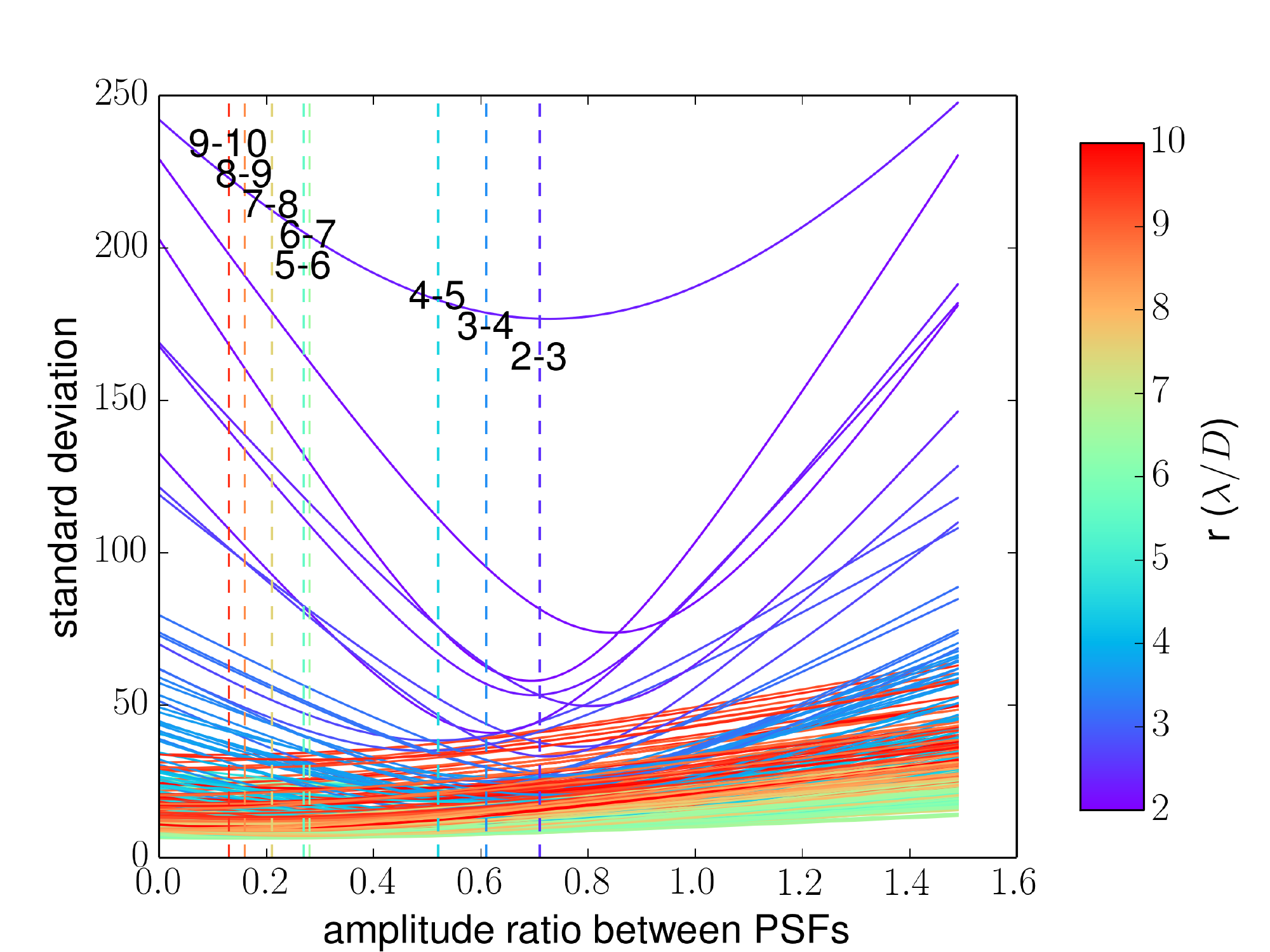}}
\caption{Standard deviation of many randomly selected pixel patches inside the combined dark holes as a
function of the scaling factor between the two PSFs.
The lines are color coded according to their distance from the star.
For different radial bins, the average scaling ratio that creates the minimal noise value is shown with the vertical lines and is labeled with the inner and outer angle (in units of $\lambda/D$) of that bin.
}
\label{fig:probes}
\end{figure}

\begin{figure}[!ht]
\center{\includegraphics[width=0.45\textwidth]{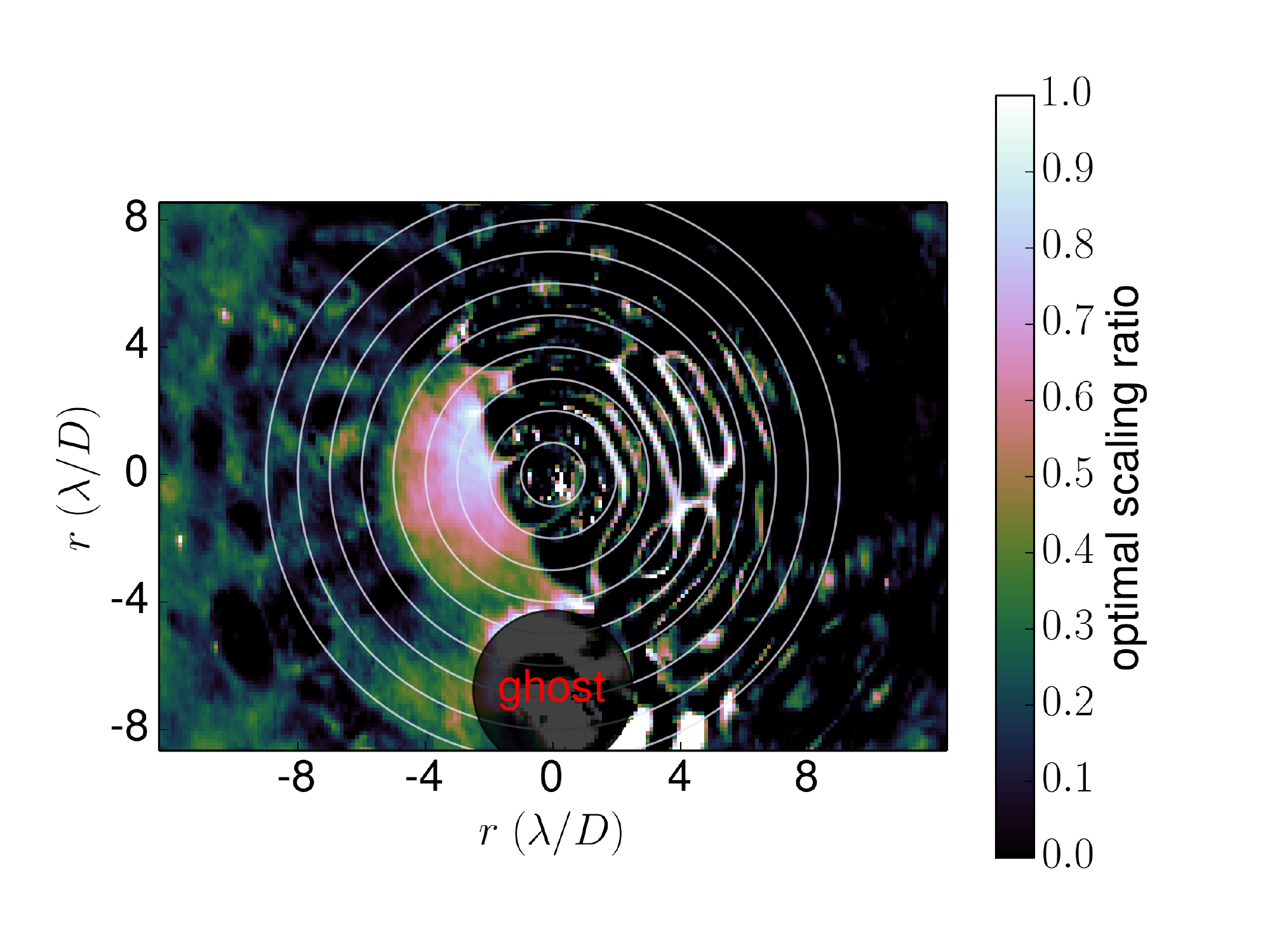}}
\caption{Map showing the optimal scaling factor for every pixel in the data cube.
Far away from the star the optimal ratio is close to zero, as the noise is fully random and uncorrelated between the two PSFs.
Close to the star in the seeing halo a ratio of about 0.7 is required for optimal noise reduction. 
}
\label{fig:ratiomap}
\end{figure}

Another option for scaling the two PSFs is to take the intensity ratio that minimizes \emph{the halo noise in time} and applying that to all frames (see also~\citet{Marois:06}).
To determine this ratio, we calculate the standard deviation for the temporal intensity variation in many
randomly selected 3$\times$3 pixel patches inside the combined dark hole.
Figure \ref{fig:probes} shows the standard deviation for various 3$\times$3 patches, which are color coded according to angular separation, as a function of the applied intensity ratio.
The vertical lines indicate the ratio at which the noise is minimal on average for a series of $\lambda/D$ bins.
As reducing the noise closest to the star is the most important, the value 0.71, which on average minimizes the noise in the bin at $2-3\ \lambda/D$, is used to scale the amplitude of the bottom PSF cube before
subtracting it from the top PSF.
Figure ~\ref{fig:ratiomap} shows the optimal scaling factor to minimize variance for each pixel inside the combination of dark holes.
It is apparent that rotation-subtracting PSFs is only effective close to the star, at the location of the seeing-driven halo.
Farther away from the star it is preferred to not perform any subtraction at
all, as at the outer parts of the dark holes the noise is uncorrelated (e.g.,~photon shot noise from the thermal background and read-noise), and therefore subtracting the two images will actually inject noise
and consequently increase it with a factor of $\approx \sqrt{2}$.
This effect is also the likely cause of the reduced optimal factor to minimize variance (0.71), in comparison to the factor of 1.04 that we found to optimally balance out the intensity structure.
A further minimization of the variance in the combined dark hole can be achieved by optimizing the ratio in radial bins as is commonly done with localized optimization of combination of images \citep[LOCI][]{lafreniere:07} and principal component analysis \citep[PCA][]{Amara:12,Soummer:12}.

\begin{figure*}[!ht]
\center{\includegraphics[width=0.9\textwidth]{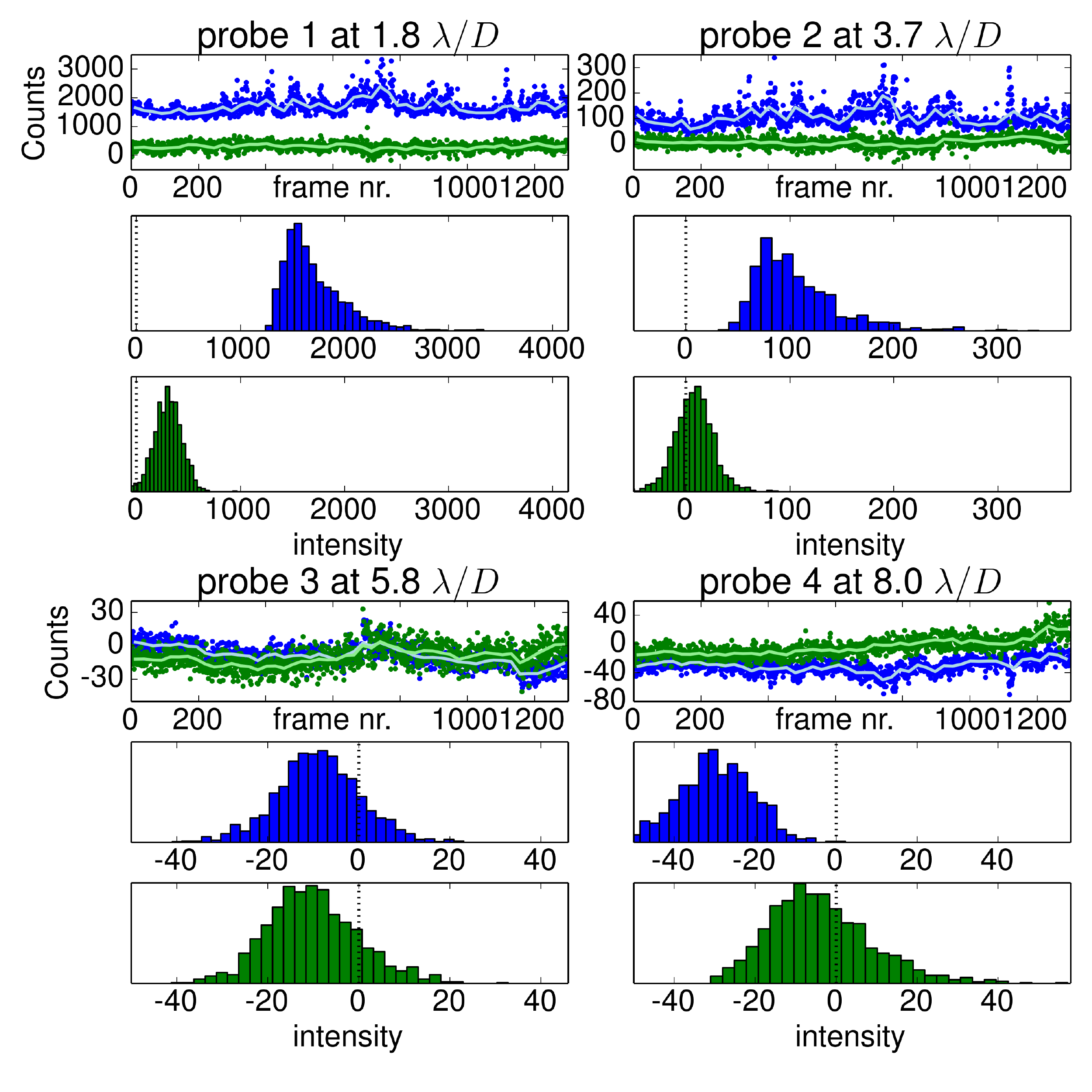}}
\caption{Time series and histograms before and after PSF subtraction for each of the four pixel patches shown in Fig.~\ref{fig:drieluik}. A boxcar-averaged line is overplotted for both cases. 
Within 4.5 $\lambda/D$ the histograms after subtraction have an average closer to 0 and become more Gaussian, and their width decreases.
}
\label{fig:allhisto}
\end{figure*}

For a scaling ratio of 0.71, we plot in Fig.~\ref{fig:allhisto} the time series and histograms for three 3$\times$3 pixel patches at four locations inside the dark holes as indicated in Fig.~\ref{fig:drieluik} before and after the subtraction to see how the rotation-subtraction technique improves the intensity variability.
At the location closest to the PSF core (1.8 $\lambda/D$), both the average value and the standard deviation of the intensity are significantly reduced. This effect is seen in both the time series and the histograms.
This is particularly evident in cases of worse AO performance (for instance, around the $\#=750$ mark).
Moreover, the rotation-subtraction technique produces histograms that are much more Gaussian than before.
As discussed already, pixels farther away from the central star obtain a $\sim$$\sqrt{2}$ increase in the noise, as their noise properties are already close to Gaussian and independent.

\section{Results: contrast curve}

The combination of the intrinsic coronagraphic performance of the vAPP coronagraph inside the dark holes and the optimal rotation-subtraction of its two complementary PSFs to subtract the residual seeing-driven halo delivers essential suppression at very small angular separations from the central star to detect and characterize planetary companions.
We apply median-filtering and classical angular differential imaging \citep[ADI][without excluding frames based on the angular distance]{Marois:06} to further suppress static and quasi-static speckles inside the combined dark holes to reach the ultimate contrast.
After rotation-subtracting the two PSFs with the optimal ratio, the median value inside a wedge for $5-7\ \lambda/D$ in the dark hole is subtracted from every pixel in every frame of the data cube. 
This process is repeated for every frame to remove any residual intensity offsets.
The median across the time dimension per pixel is removed from the whole cube to remove any residual static PSF structures.
After these steps, the frames are derotated to the sky frame and coadded by taking the mean across the time dimension.

\begin{figure}[!ht]
\center{\includegraphics[width=0.45\textwidth]{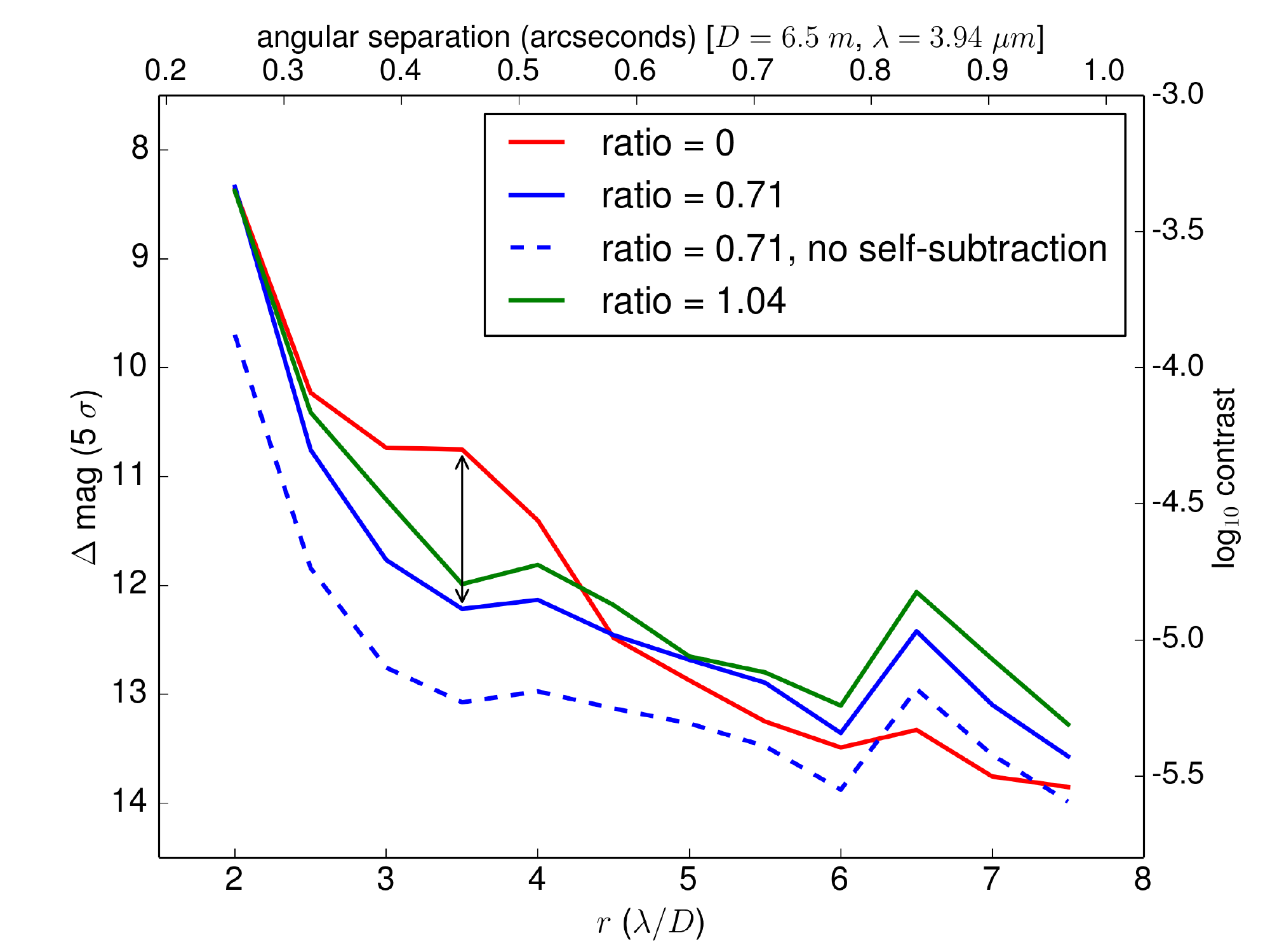}}
\caption{The 5$\sigma$ contrast curve as a function of angular separation from the central star for different scaling factors. 
The contrast is calculated after regular ADI and a mean combination of the derotated images.
The PSF rotation-subtraction improves the contrast within $4.5\ \lambda/D$ up to 1.46 magnitudes compared to using ADI in a single dark hole.
The dashed line shows the contrast if there is no self-subtraction of planet flux due to the data reduction.}
\label{fig:contrast}
\end{figure}

To assess the contrast performance, artificial companions are injected in the original data cube at steps of 0.5 $\lambda/D$ and with steps in magnitude of 1 with the expected amount of sky rotation.
The injected sources are a rescaled and translated version of the unsaturated calibration data set and therefore have the correct PSF for each dark hole.
The previously described pipeline of optimal rotation-subtraction, median-filtering, and ADI is applied to these data cubes with injected sources of varying contrast ratio.
The S/N of these planets is calculated by calculating both the sum of the flux in an aperture with a width of $1\ \lambda/D$ and the noise in the same aperture without the planet added.
The standard deviation in this aperture is multiplied by the square root of the number of pixels in the subaperture to obtain the measurement noise on the planet flux, assuming that this noise is Gaussian (which is supported by the results in Fig.~\ref{fig:allhisto}).
The magnitude of the injected point source is rescaled to obtain an $\mathrm{S/N}=5$, and these values are plotted as a contrast curve for 5$\sigma$ point source detection sensitivity versus angular separation in Fig.~\ref{fig:contrast}.
Although this method is necessarily different from the procedure introduced by \citet{Mawet:14}, as at small $\lambda/D$ the dark hole is too small to obtain a measure of the standard deviation at neighboring patches, it is fully consistent for pure Gaussian noise.
In any case, the contrast performance is clearly validated by the fact that the injected point sources at the corresponding contrast ratios are detected with large S/N, and the numbers are therefore reliable at least within a factor of a few (which is fairly insignificant on a logarithmic scale).

In Fig.~\ref{fig:contrast} we note that within $4.5\ \lambda/D$, the contrast performance is significantly improved by subtracting the other PSF with a fixed amplitude scaling factor of 0.71. 
This is most evident at a angular separation of  $3.5\ \lambda/D$, where the improvement is 1.46 magnitudes (four-fold improvement) to a $\Delta \mathrm{mag}$ of 12.2, which corresponds to a contrast of $10^{-4.8}$.
Beyond $4.5\ \lambda/D$, the contrast performance for the rotation-subtraction technique is degraded, as here the noise is random and uncorrelated, and therefore aggravated after the combination with the second PSF. 
As previously mentioned, we expect to be able to reduce this effect by optimizing the scaling factor in radial bins, although this also increases the degrees of freedom. 
The turnover point at $4.5\ \lambda/D$ is dependent on the brightness of the target as it moves inward with fainter targets as the background noise contribution becomes more dominant.
For this data set, the turnover point at $4.5\ \lambda/D$ has a $\Delta \mathrm{mag}$ of 12.5, corresponding to a 5$\sigma$ contrast of $10^{-5}$.
Like many other reduction methods, our classical ADI approach also removes part of the planet flux in addition to residual speckles in the stellar PSF.
To quantify this effect, we retrieve the planet flux after applying the entire data-reduction pipeline to the data and compare it to the injected planets.
The efficiency of the ADI algorithm is as low as 29\% at 2 $\lambda/D$ and reaches 68\% at 7.5 $\lambda/D$.
This lower efficiency close to the star is expected as there is less angular displacement of the planet in terms of $\lambda/D$ which leads to more self-subtraction. 
We overplot in Fig.~\ref{fig:contrast} the 5$\sigma$ contrast excluding self-subtraction, which would reach down to below $10^{-5}$ for $>$3 $\lambda/D$.
This limiting case may be reached by applying more advanced PSF subtraction techniques, like Principal Component Analysis \citep{Amara:12,Soummer:12}.

\section{Discussion and conclusions}
\label{sec:disc}

To put this contrast performance of the vAPP coronagraph at MagAO/Clio2 in context, we compare our results to published on-sky contrast curves for different coronagraphic instruments.
Such an analysis necessarily uses heterogeneous data sets because of variations in the brightness of the star, the wavelength, the size of the telescope, and the applied data-reduction techniques.
%
%
It is important to note, however, that close to the star the brightness of the star has little impact as the contrast there is limited by speckle halo noise.
Moreover, all published contrast curves are produced for stars that are bright enough that the AO system still has its optimal performance.
Furthermore, we put all curves on a $\lambda/D$ scale to account for differences in telescope diameter and observation wavelength, which provides the most honest comparison.

To begin with a related coronagraph, in comparison to the performance of the regular APP at VLT/NACO \citep{Quanz:10,Kenworthy:13,Meshkat:14}, the vAPP PSFs do not exhibit any clear diffraction structure close to the star, whereas the VLT APP PSF clearly does.
The much-improved manufacturing accuracy of the phase patterns now permits the creation of dark holes that are devoid of diffraction structure down to $10^{-5}$.
Moreover, the coronagraphic PSFs of the grating-vAPP are not deteriorated by leakage PSFs, as they form a separate PSF, which actually can be used to one's advantage as a photometric or astrometric reference.

\begin{figure}[!ht]
\center{\includegraphics[width=0.45\textwidth]{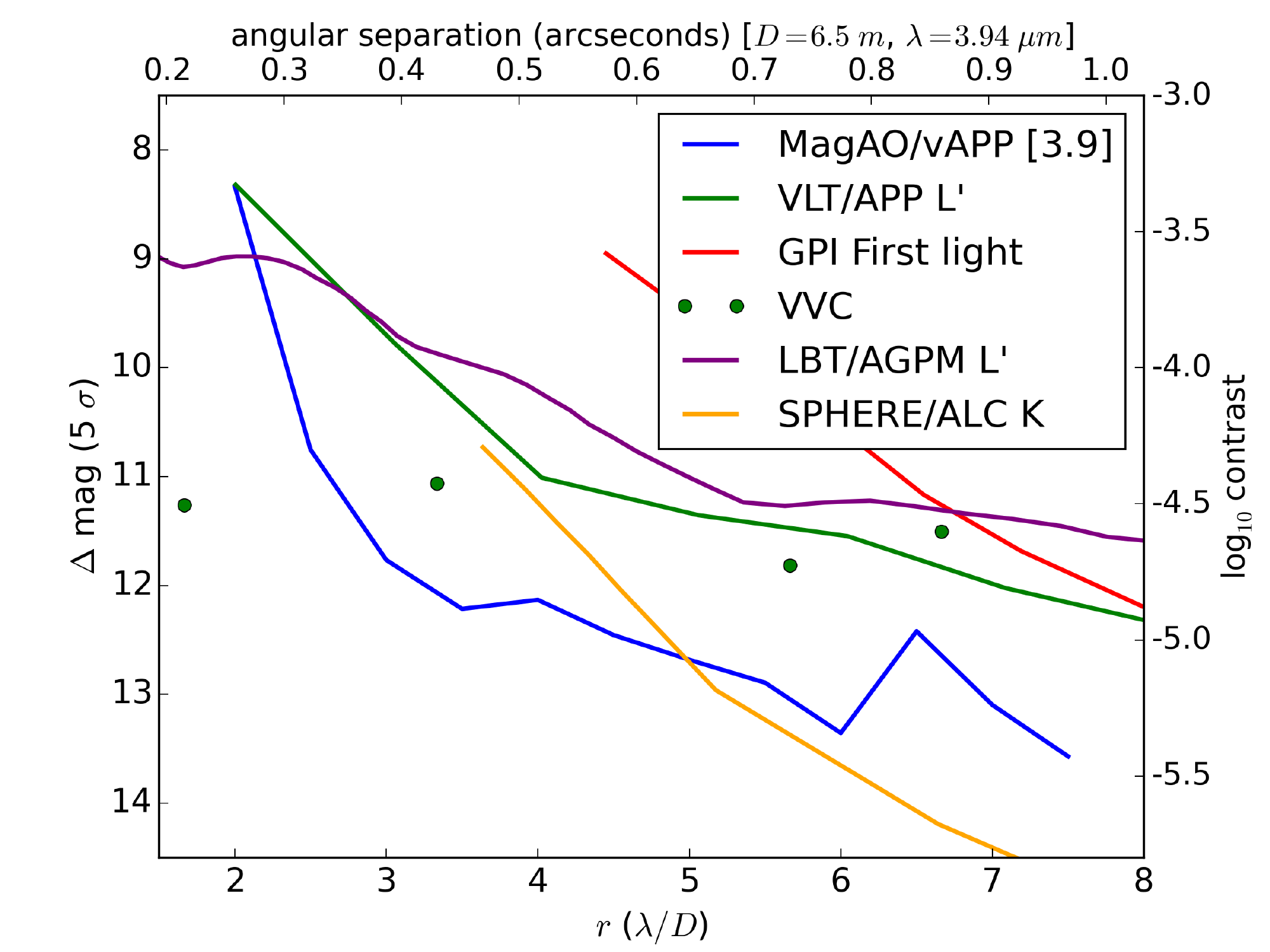}}
\caption{Comparison of $5\,\sigma$ contrast as a function of angular separation from the central star from the literature. The published contrast curves of other studies have been overplotted at the same $\lambda/D$ to correct for different telescope sizes and observing wavelengths. The vAPP outperforms many other coronagraphs close to bright stars where one expects to be speckle and tip/tilt limited.}
\label{fig:comparecontrast}
\end{figure}

The MagAO/Clio2 gvAPP coronagraph contrast performance from Fig.~\ref{fig:contrast} is compared in Fig. \ref{fig:comparecontrast} with the following contrast curves: the annular groove phase mask (AGPM) at LBT \citep{Defrere:14}, the vector-vortex coronagraph (VVC) at the 1.5 m well-corrected aperture at Palomar \citep{Serabyn:10}, the GPI first-light results \citep{Macintosh:14}, SPHERE with the apodized Lyot coronagraph (ALC) \citep{Vigan:15}, and the APP at the VLT \citep{Meshkat:14}. 
All these published contrast curves are corrected from the published $N\sigma$ to a $5\sigma$ detection limit. 
We assume the contrast is not limited by photon noise in all cases, and therefore we do not correct the curves for differences in exposure time and telescope diameter. 
In terms of $\lambda/D$, both the GPI and SPHERE contrast curves tend to reach high contrasts farther away from the star, which is likely due to the fact that they were taken at shorter wavelengths where the sky background is lower than in the $L$ band. 
Most notably, the vAPP has a much smaller inner working angle (IWA) in combination with better contrast performance at small angular separations than GPI and SPHERE, when measured in $\lambda/D$. 
The IWA of the SPHERE ALC is restricted by the focal-plane mask to 120 mas.
Moreover, all focal-plane coronagraphs are limited in their contrast performance at small $\lambda/D$ due to imperfect tip/correction.
The VVC result at the Palomar 1.5 m well-corrected aperture is a bit of an outlier because of the significantly different $D/r_0$ ratio as compared with the other telescopes, but it is included because of its high performance at a $\lambda/D$-sized IWA. 
The assumption of being speckle limited likely does not hold here as the VVC results were within a factor of two from the photon noise on the background.
Nevertheless, we see that the vAPP is a very strong contender or even outperforms the other coronagraphs within $5\ \lambda/D$ with an improvement of up to 2 magnitudes for $2.5-3.5\ \lambda/D$.

The exceptional contrast performance of the vAPP coronagraph is owed to the unique combination of the following properties:
\begin{enumerate} 
\item Insensitivity to tip/tilt errors that impact focal-plane coronagraphs but not pupil-plane coronagraphs like the vAPP.
\item Deep suppression of the PSF diffraction structure with an accurately manufactured (geometric) phase pattern already at the first diffraction ring down below the seeing-driven halo.
\item Subtraction of the halo in the dark holes by combining both PSFs with a rotation-subtraction technique.
\end{enumerate}
We see that using the second coronagraphic PSF as a PSF reference gives an improvement of $1-1.5$ magnitudes (a factor $2.5-4$ in terms of S/N) at $3-3.5\ \lambda/D$. 
The PSF subtraction is shown to improve the contrast within $4.5\ \lambda/D$. 
With a radially optimized subtraction, the degradation of the contrast outside this distance can be reduced.
Given a fixed ratio based on optimal contrast close to the star, we achieve a $5\sigma\ \Delta \mathrm{mag}$ contrast of 10.8 ($=10^{-4.3}$) at 2.5 $\lambda/D$, 12.2 ($=10^{-4.8}$) at $3.5\ \lambda/D$, and 12.5 ($=10^{-5.0}$) at $4.5\ \lambda/D$.
Using a PCA-based algorithm instead of applying classical ADI, we expect that our performance will be less impacted by self-subtraction and will improve toward the dashed line of Fig. \ref{fig:contrast}. 
Use of a simultaneous reference PSF was also explored by \citet{Dou:15}, who used the roughly symmetric PSF itself under rotation to feed a PCA algorithm. Their approach gave an improvement of an order of magnitude in terms of contrast when compared to LOCI.
\citet{Rodigas:15b} used a close binary star to build their reference PSF library.
By having a simultaneous reference within the isoplanatic patch, and with roughly the same optical path through the telescope, a better sensitivity is expected than using the star as its own reference. In their study, a 0.5 magnitude improvement within 1 arcsec from the star was seen as compared to normal ADI.
\citet{Rodigas:15b} suggest combining their binary differential imaging (BDI) technique with the vAPP coronagraph to reach better contrasts.
We can extend this by noting that a double correction can be done by combining BDI and the second vAPP PSF as another reference.
In both previous cases, the methods are less impacted by self-subtraction because it is unlikely that a companion exists in the reference library with similar brightness, position angle, and separation.
Both papers give us confidence that an advanced PCA-based algorithm can be used to generate a better reference PSF and the contrast can be pushed down even more. 

To improve the transmission of the optics, one of the three substrates and consequently one adhesive layer could be eliminated by directly depositing the liquid crystal layer on top of the antireflection-coated substrate and bonding it directly with the substrate with the aluminum mask.
This procedure increases the transmission by about 20\% but makes manufacturing slightly more difficult and expensive.

\subsection{Future work}

We have demonstrated a manufacturing technique, based on the geometric phase imposed by to patterned liquid crystals, that allows precise control of the phase pattern and a broadband coronagraphic response that can be optimized at any wavelength range from the UV to the mid-IR. 
The gvAPP coronagraph is relatively straightforward to manufacture and install at existing telescopes as it consists of a single optic in a pupil plane. 
Using these capabilities, we are looking into new phase patterns with properties different from that normally expected from classical APP theory.
For instance, a vAPP with a dark hole spanning $360^\circ$ per PSF or with integrated holographic wavefront sensing solutions \citep{Wilby:2016} has been implemented and tested on sky.
Future work also includes exploring hybrid coronagraphs, for instance as described by \citet{Ruane:15}.
Furthermore, we will study the photometric and astrometric stability of the leakage term to assess how well this works as a reference PSF.
Following the implementations described by \citet{Snik:14spie}, we also intend to explore the dual-beam polarimetric capabilities of the vector APP in the optical lab and on sky. 
For polarized sources, an increased sensitivity is expected by simultaneously using the coronagraphic capabilities and polarimetric beam-switching.

\section*{Acknowledgments}
\noindent We thank the anonymous reviewer for constructive comments that helped improve this paper. This work is part of the research program Instrumentation for the E-ELT, which is partly financed by the Netherlands Organization for Scientific Research (NWO). The MagAO vAPPs were purchased from ImagineOptix with help from the Lucas Foundation and the NASA Origins of Solar Systems program. FS is supported by European Research Council Starting Grant 678194 (FALCONER). KMM’s and LMC’s work is supported by the NASA Exoplanets Research Program (XRP) by cooperative agreement NNX16AD44G.
This research has made use of the SIMBAD database, operated at CDS, Strasbourg, France. This paper includes data gathered with the 6.5 meter \textit{Magellan} Telescopes located at Las Campanas Observatory, Chile.

\facility{Magellan:Clay (MagAO/Clio2)}.

\bibliography{apj_otten_proof}
\bibliographystyle{apj}

\end{document}